\documentclass[twocolumn,showpacs,preprintnumbers,superscriptaddress,amsmath,amssymb,prl]{revtex4}

\usepackage{graphicx}% Include figure files
\usepackage{dcolumn}% Align table columns on decimal point
\usepackage{bm}% bold math
\usepackage{amsmath}

\newcommand{\be}{\begin{equation}}
\newcommand{\ee}{\end{equation}}
\newcommand{\ba}{\begin{eqnarray}}
\newcommand{\ea}{\end{eqnarray}}
\begin{document}

\preprint{APS preprint}

{\bf Intercluster Correlation in Seismicity}

\begin{figure}
\includegraphics[width=8cm]{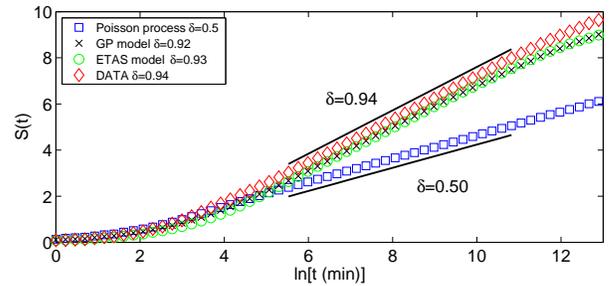}
\caption{\label{Fig1} The DE $S(t)$ defined in \cite{Mega} calculated for
a Poisson process (squares), the California catalog used in \cite{Mega} (diamonds),
the GP model used in \cite{Mega} and the ETAS model (circles) with 
parameters: branching ratio $n=0.93, b=1$, Omori exponent $p=1.3$
for first-generation triggering and $\alpha=0.8$.
}
\end{figure}

\begin{figure}
\includegraphics[width=8cm]{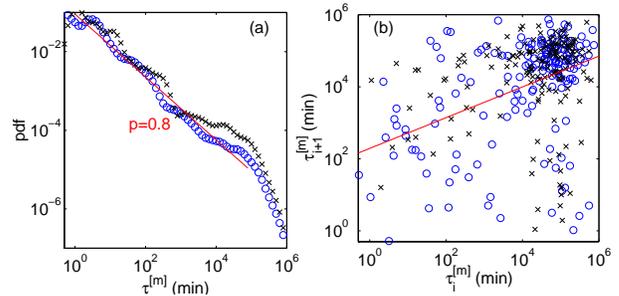}
\caption{\label{Fig2} (a) probability density function (pdf) of the time 
intervals $\tau^{[m]}$ between two successive large ($M\geq 5$) earthquakes
 in the Southern California catalog (circles) and in the ETAS model (crosses). 
(b) recurrence plot $\ln \tau^{[m]}_{i+1}$ vs $\ln \tau^{[m]}_{i}$.
The straight line is the linear fit to the data (circles). The correlation is 
$r=0.43$ (significance $>99.9\%$) for the data and $r=0.21$
(significance $>99\%$) for ETAS.
}
\end{figure}

\noindent Mega et al. \cite{Mega} proposed to use the 
``diffusion entropy'' (DE) method to demonstrate that
the distribution of time intervals between a large
earthquake (the mainshock of a given seismic sequence) and the next one
does not obey Poisson statistics.
We have re-analyzed the same catalog of seismic events in California
and performed synthetic tests which 
show that the DE is unable to detect correlations between clusters,
thus negating the claimed possibility of detecting an intercluster correlation.
We have generated synthetic earthquake catalogs with a Poisson
distribution of mainshock times, with aftershocks triggered 
in direct lineage by each mainshock following  Omori's power-law decay 
of seismic rate $n(t) \propto 1/(t-t_i)^p$
(with $p=1$, and where $t_i$ is the mainshock time) 
and with a power law distribution of aftershock cluster sizes with exponent 
$b/\alpha=1.25$ (justified from the interplay between the power-law
distribution of energies with exponent $b=1$ (Gutenberg-Richter law) 
and the power-law increase of aftershock productivity with the
mainshock energy of exponent $\alpha=0.8$ \cite{alpha}).
Fig. \ref{Fig1} (crosses) shows that our synthetic DE without intercluster 
correlation is essentially undistinguishable from the real data.
Thus, the conclusion of Mega et al. \cite{Mega} that the clusters of
seismicity are correlated is not warranted by their analysis.
However, the {\it fact} that intercluster correlation exists is known
at least since Kagan and Jackson \cite{KJ} and can be observed straightforwardly, 
as shown in Fig. \ref{Fig2}. 

We have used the ETAS model to generate realistic synthetic catalogs.
The ETAS model is a simple branching model of seismicity \cite{Ogata}
which contains the Gutenberg-Richter, the Omori law, the productivity law, 
cascades of multiple triggering between earthquakes and a poissonian
seismicity background. In the ETAS model, any earthquake may trigger
other earthquakes, without arbitrary distinction between foreshocks,
aftershocks and mainshocks. The ETAS model reproduces
many properties of seismicity, including realistic foreshock sequences
 \cite{foreemp}. It is widely used to model and predict
the spatio-temporal distribution of seismicity  (e.g., \cite{Ogata} and
ref. 8 of \cite{Mega}).
%such as 
%the larger proportion than normal of large versus small foreshocks,
%the power law acceleration of seismicity rate as a function of
%time to the mainshock and the spatial migration of foreshocks toward 
%the mainshock, when averaging over many sequences \cite{foreemp}.
As shown in Figs. 1 and 2 (circles), the ETAS model recovers
(i) Mega et al.'s observation of DE $S(t) = A + \delta \ln  t$ with
$\delta = 0.94$, (ii) the power-law pdf
of the time  intervals $\tau^{[m]}$ between two successive mainshocks
(which arises because many mainshocks are also aftershocks of other mainshocks) and
(iii) a correlation between $\tau^{[m]}_i$'s (which is absent in Mega et al.'s LR model).
Thus, their LR model, which introduces correlation between clusters of seismicity, is
insufficient to account for the correlation observed in the data.

We believe that these discrepancies with Mega et al.'s conclusions stem for their
incorrect use of the pdf of intracluster interevent times $t_{j+1}-t_j$. Omori's law 
is the pdf of the times between a mainshock and its aftershocks and describes a
non-stationary process such that the interevent times increase as time increases
since the mainshock: for $p=1$ for instance, $t_{j+1}-t_j \sim t_j$.
Therefore, the pdf of interevent times is not equivalent to Omori's law and it
looses most of the information on correlations between aftershocks.
This work is supported by NSF-EAR02-30429 and by
the Southern California Earthquake Center (SCEC).

A. Helmstetter$^1$ and D. Sornette$^{1,2}$,
$^1$ Institute of  Geophysics and Planetary Physics, UCLA
$^2$ Department of Earth and Space Sciences,
University of California, Los Angeles, California 90095-1567
and LPMC, CNRS UMR 6622 and
Universit\'{e} de Nice, 06108 Nice, France

\end{document}